\documentclass{scrartcl}

\usepackage[]{graphicx} 
\usepackage[utf8]{inputenc}
\usepackage[T1]{fontenc}
\usepackage{lmodern}
\usepackage[english]{babel}
\usepackage{amsmath}
\newcommand{\f}{\footnote}
\usepackage{xspace}
\usepackage{natbib}
\expandafter\def\expandafter\grqq\expandafter{\grqq\xspace}
\usepackage{amssymb,amsthm}
\usepackage{thmtools}
\usepackage{color}

\begin{document}
\title{Criticizing Ethics According  \\ to Artificial Intelligence}
\author{Irina Spiegel}
\maketitle

\textbf {Abstract}
\begin{small} This article presents a critique of ethics in the context of artificial intelligence (AI). It argues for the need to question established patterns of thought and traditional authorities, including core concepts such as autonomy, morality, and ethics. These concepts are increasingly inadequate to deal with the complexities introduced by emerging AI and autonomous agents. This critique has several key components: clarifying conceptual ambiguities, honestly addressing epistemic issues, and thoroughly exploring fundamental normative problems. The ultimate goal is to reevaluate and possibly redefine some traditional ethical concepts to better address the challenges posed by AI.
\end{small}
\tableofcontents

\section{Preliminary notes}
The project of a new critique aims to systematically analyze epistemic and ethical concepts and patterns of thought commonly used today in connection with digitalization, artificial intelligence, and big data. These concepts are often employed without sufficient scrutiny. The central thesis posits that this lack of critical examination—whether self-inflicted or not—prevents achieving genuine problem-solving orientations in areas increasingly dominated by autonomous AI systems, such as care robots in nursing homes, surgical robots in medicine, and war drones. However, the proposed new critique does not primarily focus on the concrete ethical problems within the specific fields of AI application. Instead, it addresses the fundamental epistemic and normative questions that arise in relation to future artificial intelligence (AI). These questions demand thorough examination to better understand and navigate the ethical landscape shaped by advancing AI technologies. The discussion centers on the conditions under which a future AI can be normatively and/or ethically\f{In this context, ``normative'' refers to a set of top-down, principle-based necessary conditions of normativity. Conversely, ``ethical'' pertains to a set of bottom-up values and norms based on a society’s environment.} embedded. The urgency of this project is clearly demonstrated by the latest AI revolution, exemplified by Chat GPT.
\\
\\
There are roughly two levels of critique that can be identified. The level of trivial critique simply involves a broad, pervasive, and comprehensible dissemination of information to citizens about the available means to protect their privacy (e.g., using Startpage instead of Google), about data sovereignty, about the problem of the loss of privacy, about platform capitalism, and the neo-feudal power expansion of tech monopolies, etc. All of these topics have already been discussed relatively extensively, but digital literacy has not been widely taught, even though it could be implemented relatively easily and directly\f{e.g., as an information leaflet included with election notices for local, state, or federal elections.}. This kind of trivial critique has also not been sufficiently integrated into today’s educational systems. A comprehensive political and educational agenda would be needed here.
\\
\\
A deeper level of new critique involves a fundamental problematization, as this project challenges and radically questions cherished ethical and epistemic thoughts. This questioning is necessary because of the accumulation of unresolved normative and ethical issues in AI applications, and the suspicion that this is primarily due to traditional, established ethical concepts and beliefs. The corresponding ethical discourses are circular and reveal unresolved normative fundamental problems. It takes courage to imagine how future AI will ethically challenge our conception of humanity and the world. And it takes courage to admit that established ethical practices, beliefs, and theories are limited, and therefore need not only be questioned, but also developed. Normative ethics in general, and ethical or humanistic intuitions in particular, must themselves be subjected to scrutiny.
\\
\\
The underlying assumption of the deep critique is that contemporary philosophical ethics in general cannot keep up with the rapid developments in AI. It is unrealistic to expect philosophers and ethicists to have sufficient expertise in AI, and the interdisciplinary exchange between computer scientists and ethicists or philosophers is proving to be challenging. In addition, programmers and developers increasingly do not understand why and how exactly their algorithms produce certain results (the black-box problem). Although there is more or less willingness to work across disciplines, the concepts of ethics and computer science are not congruent, which hinders interdisciplinary exchange. This seems to be partly due to the nature of AI and partly due to the traditional concepts of ethics themselves.
\\
\\
In contemporary philosophical ethics, there is an uneasy coexistence of different approaches. However, it has become clear that any attempt at unification or mediation (e.g. between deontology and consequentialism) leads to contradictions. This fundamental normative incoherence is also reflected in the application areas of digital ethics and leads to inconsistencies in the potential implementation of ethical concepts. For example, for a rescue robot, the number of human lives is both operationally decisive (consequentialist) and not decisive, since from a deontological perspective, everyone has the right to an equal chance of being rescued, regardless of whether they are in a large or small group, or whether an individual is weighed against a group. Normative and applied ethics are particularly limited in the field of digital technology and AI, as normative rules and principles would have to be programmed in here.
%
\\
\\
This project begins by analyzing the notorious conceptual and semantic ambiguities involved. Philosophy and ethics, computer science, AI, futuristic visions, and science fiction often use identical terms, although sometimes with different meanings and connotations. In relation to specific AI applications, the ambiguity of terms such as ``intelligence'' and ``autonomy'' can fuel fears as well as hopes and lead to corresponding individual or collective behaviors in the real world (dystopian apathy or euphoric complacency). These conceptual ambiguities not only mislead the general public but also cause people to misjudge the risks and potentials of AI or to be unable to refrain from making such presumptuous assessments. 
\\
\\
The moral, legal, and political questions and concerns associated with contemporary AI are increasingly occupying the attention of the public, ethicists, and politicians. There is a certain consensus in these discourses that an autonomously acting AI should ideally not be regulated after the fact and should be normatively constituted before implementation, given that certain developments in this field are irreversible. In reality, however, post-hoc regulation remains the norm in the field of AI, as the recent debates around ChatGPT have shown. The necessary sluggishness of ethical and political debate and decision-making contrasts sharply with the frenetic pace of AI development. This has led even those directly involved in AI development (e.g., Elon Musk and Bill Gates) to call for a slowdown in AI progress. This is intended to buy time to carefully consider how intelligent (autonomous) systems might be ethically, politically, and legally constrained and regulated. This pause in AI development should be used to explore the possibilities of ``normative programming'' or a ``moral machine'' (Wallach 2008), so that future AI conforms to ethical and democratic rules and norms right from the design stage. However, current political and ethical debates and attempts at normative or moral programming reveal intractable normative problems, paradoxes, and dilemmas. It seems that without a questioning of traditional normative and ethical concepts and without new theoretical normative approaches, the ethical issues of AI will not be manageable. This is a negative thesis of the deep critique. How can humans implement values and norms in machines if they themselves do not have a consistent normative theory that can be systematically applied for the humanization of the world?
\\
\\
In addition to the \emph{normative} problems, there is an \emph{epistemic} challenge regarding the explainability and interpretability of AI systems, their algorithms, and the results they produce (i.e. the field of Explainable Artificial Intelligence). Ethical decisions often depend on specific distinctions or object classifications. However, with existing deep learning software, it is increasingly difficult for humans to understand how the system arrived at a solution or classification. This makes the decision-making processes of such AI systems uncontrollable. The opacity of the inner workings and reasoning of advanced machine learning models exacerbates the difficulty of aligning their behavior with ethical principles. When the logic underlying AI decisions cannot be adequately explained or understood, it becomes extremely difficult to ensure that those decisions are ethically sound, legally compliant, and socially acceptable.
\\
\\
The problem of classificatory distinction suggests that knowledge generated by AI will challenge, or is already challenging, current notions of knowledge and explanation. The volume of (statistical) knowledge and correlations is growing rapidly with ever larger big data. This leads to a very large number of new correlations and probabilities that would have to be accompanied by ``explanation'' (and thus ``understandin''”) or would remain without it. Many AI applications depend directly on probabilistic knowledge, and it is unclear whether humans have any real ``knowledge'' in these areas. For example, computational data science could recommend interventions or recommendations for health care and environmental protection that would be difficult for humans to understand. In this context, is it possible to develop a fundamental trust in AI systems? According to Yuval Harari, it depends on people’s experience with these AI recommendations, whether AI statistically much often deliver very good results or much better results than human experts (Harari 2018).
\\
\\
How should a future ethics of AI be conceived in this context? It is likely to be less about purely technical progress and more about the need for new ideas of ``social or normative progress''. By social or normative progress, I do not mean that the ethical problems of AI applications can simply be solved by appropriate laws and regulations in states plagued by multiple crises today. Instead, new normative ideas and theories are needed.
More advanced AI technology alone will not solve humanity's problems. Solving them will probably require surprising ideas for a new global, societal and political reorganization, which will derive legitimizing (normative) power from the practical application of a new normative theory. People will have to learn to see AI as a kind of tool or even ``partner'' in a social redesign based on the foundations of a new normative theory.
\\
\\
This dimension is almost entirely absent from current ethics. It still consists primarily of disparate ethical discourses related to the concrete manifestations of contemporary AI, which consistently show a strong connection to established ethical paradigms, especially traditional deontology and consequentialism. However, in order to address the fundamental problems and challenges of AI in an adequate and constructive way, this methodology - the application of established ethics - seems insufficient. This inadequacy stems 
from the radically transformative nature of AI technologies themselves.
\\
\\
The project of a deep  critique aims to add a critical meta-level to the debates in current AI ethics, and to question and analyse the way ethics (along with its philosophical entanglements) is practiced today. What can it still achieve in its established form, and what cannot? What problems arise from current approaches to AI ethics? And how could and should a new, more advanced ethics and AI ethics be constituted? These are some of the core questions of a deep critique of ethics as a work in progress.
\\
\\
In order to develop a better sense of the challenging ethical and theoretical problems that may await us, four essential processes of AI elucidation are outlined below. These are clarifying conceptual ambiguities, critically reflecting on AI risks, considering epistemic issues, and examining fundamental normative problems.

\section{Clarifying conceptual ambiguities}
In AI research and ethics, the terms ``intelligence'', ``agency'' and ``autonomy'' are regularly used. But computer scientists and philosophers often understand these terms differently, sometimes even completely differently.

\subsection{Intelligence}
Since philosophy began,``intelligence'' has come to be seen as a specific characteristic of human beings. It has come to include non-human forms of intelligence as well as machine intelligence. This concept is central to both AI research and public discussion. It is therefore necessary to redefine and reinterpret its meanings.
\\
\\
In modern philosophy, the term 'intelligence' was nearly synonymous with the concept of reason, understood as the human capacity to comprehend the regularities of the external and internal worlds. However, the definition of intelligence is shifting with the emergence of new approaches that consider it a collection of cognitive abilities that can be quantified and modeled. These developments have arisen from a convergence of disciplines, including the natural and social sciences, as well as psychology. The concept of intelligence in AI research is based on this scientific notion of intelligence as a set of mathematically describable cognitive functions (cf. McCarthy et al. 1955), which can be implemented in AI and embodied by robots.
\\
\\
Currently, artificial intelligence (AI) is increasingly being attributed\f{This phenomenon can be attributed, in part, to the prevalence of popular science fiction narratives.} with qualities such as emotional perception, reflective capacity, and consciousness (see, e.g., Misselhorn 2018). In this context, AI ethicists are also consulting so-called philosophy of mind more frequently (ibid.). Here, ``intelligence'' is also coupled with (self-)consciousness. This coupling draws a sharp boundary between the intelligence of sentient beings, especially humans, and artificial, merely algorithmic ``intelligence'', which operates mechanistically without experience. Whether this sharp boundary truly exists is questionable at best, as long as it remains unclear what exactly consciousness or experience is, which is supposed to mark and define this boundary.
\\
\\
In the philosophy of mind, the problem of consciousness is called the ``hard problem'' (Chalmers 1995). This implies that consciousness cannot be reduced to a scientific or physical theory. Nevertheless, this does not preclude the possibility that such an inexplicable, physically indescribable property could emerge in future intelligent systems or robots. To date, there is no evidence to suggest that an AI could develop a first-person perspective or have ``states'' equivalent to ``mental states'' (qualia). The majority of philosophers reject this possibility, although they often fail to provide compelling arguments. Those engaged in the serious study of computers refrain from taking a position on this matter. It remains unclear whether an AI with global access to relevant information (\emph{internal global workspace}) and capable of self-observation or meta-cognition (\emph{reflexive representation}; cf. Dehaene et al. 2017)) could possess its own peculiar mental states\f{The question ``What is it like to be a robot?'' can be seen to be analogous to Nagel's famous question ``What is it like to be a bat?''.}.

\subsection{Agency}

``Agency'' has also been a focus of modern AI research since the 1980s.\f{This English expression has become established in the literature; in German, it is called ``Handlungsfähigkeit'', which is more metaphysically charged than agency because the latter presupposes reason and freedom of will.} Agents are distinguished by their capacity for intelligent behavior, whereby they receive perceptions from the environment and are able to actively influence it (cf. Stuart J. Russell/Peter Norvig 2010). This approach is based on the classic theories of Human Problem Solving by Newell and Simon (1972): ``Every such agent implements a function that maps percept sequences to actions''. (Russell/Norvig 2010, 2) In accordance with this operational definition of an AI, the action of an AI is mapped from sensory input to mechanical output. It does not contain any goals of its own; rather, these are defined externally. In the context of philosophy, an agent is typically understood to have intentionality towards its actions (also known as an ``actor''). This implies that agents possess intentions and self-determined goals. However, for an agent to act in a philosophical sense, it is essential that the agent has \emph{deliberated} upon these intentions beforehand. This implies that an agent must have considered these intentions at some point in the past. Consequently, philosophers maintain that artificially intelligent agents cannot be considered to possess agency until they are capable of deliberating and evaluating their own self-determined goals.
\\
\\
The distinction between these two perspectives on agency – the technical (behaviorist or mechanistic) and the philosophical (intentional) – has significant ethical implications. According to most philosophers, AI agents lack the capacity to bear moral responsibilities or be held accountable for their behavior, as they have no real goals and thus cannot be expected to explain or justify their actions. The situation is distinctly different in the case of humans. However, is this truly an absolute distinction? For instance, Daniel Dennett characterizes computers as ``intentional systems'', which can be ascribed a certain purposefulness and thus regarded as ``intentional agents'' (1987). As Dennett defines it, the\emph{ intentional stance} is a prerequisite for the \emph{moral stance}. Consequently, the two states are not equivalent. According to Dennett, a moral agent must be an intentional system. However, an intentional system, such as an AI, is not necessarily a moral agent. In this context, it is possible to regard artificial agents as intentional agents without attributing moral responsibility to them. Nevertheless, they could be considered to have a similar moral status to infants and complex sentient animals, which are also considered to be intentional but not moral agents. Should an AI display behavior that is strikingly human-like or already animal-like, this could potentially challenge the prevailing mechanistic view of AI agency. What impact would this have on human agency in general and specifically in everyday interactions between humans, between humans and AI, and between robots themselves?

\subsection{Autonomy}
In AI theory and application, the adjective ``autonomous'' refers to systems and machines that can function properly without direct human intervention. The collection of information via sensors and the resulting actions occur without external input. In this sense, AI researchers speak of (more or less completely) autonomous vehicles or autonomous weapons.
\\
\\
Philosophers have criticized this usage, arguing that it blurs the boundaries between autonomy and automaticity (cf. Nida-Rümelin 2018, Gabriel 2018). In both cases, we are talking about behavior ``out of itself'', or in Greek \emph{auto}. However, the words ``autonomy'' and ``automaton'' have different connotations. Autonomy is associated with \emph{nomos}, or law, while automaton is linked to \emph{matos}, or movement. Accordingly, philosophers differentiate between beings that are strictly autonomous, which are capable of prescribing the rules or laws that govern their actions, and automata, which are externally programmed to adhere to specific movements.
\\
\\
In numerous philosophical traditions and cultures, agency and autonomy are two interrelated characteristics of responsible persons who determine their own actions in the world and can therefore also be held responsible for and must justify their actions. In the context of AI technology, an agent is defined as a software program, a computer system, or a robot that is capable of executing the commands of programmers, users, or other higher-level programs. The issue arises from the fact that learning algorithms function quite independently of programmers or users, and are more autonomous than mere automatons, such as coffee machines. It would appear that there is no absolute boundary between autonomy and automatism, and thus we are compelled to regard the boundary as fluid in this context as well.
\\
\\
It is uncertain whether, in the future, software agents may become autonomous in the sense of being responsible. At present, there is no evidence that such an AI exists, and it is challenging to make a prediction. Nevertheless, the use of autonomous, learning AI systems is already giving rise to questions regarding the attribution of responsibility that are difficult to answer. To address these questions effectively, it is essential to gain a nuanced understanding of the underlying problem and to critically examine the key concepts at play, including those of agency and autonomy. Such questioning has not yet occurred to a sufficient extent. This is probably due to the excessive authority of generally conservative philosophy regarding the interpretation of such key concepts. The terms in question are normatively charged. A fundamental normative question is what should and should not be considered autonomous. Moreover, fundamental normative questions remain unanswered, and people experience these unanswered questions as discomfort. Philosophers and ethicists today appear to address the symptoms of this unease primarily through traditional ``answers'' (of Aristotelianism, Kantianism, utilitarianism, etc.). Who will address the unresolved fundamental normative issues in the context of ongoing digitalization and the further development of AI?

\section{Critical Reflection on AI Risks}

The insufficiently clarified conceptual ambiguities, in combination with the network effects of social media and the spread of conspiracy myths, mean that AI risks are not (or cannot be) correctly assessed. On the one hand, the dangers posed by AI are overestimated, such as the fear that an autonomous AI that sets its own goals and is equipped with weapons could wipe out the entire human race. On the other hand, the very real dangers posed by the algorithms already implemented are massively underestimated today. These include the effects of the dissolution of privacy, the spread of deep fakes, and the social consequences of social scoring.

\subsection{Overestimating the Risks of AI}

In the contemporary era, there is a profusion of dystopian narratives. For example, there is concern that AI may be used to deceive humans, that humans may become cyborgs, and that AI may eventually surpass humans to an unimaginable extent, potentially ``playing God''. The existential fears associated with the advent of new disruptive technologies and their philosophical scrutiny have a long history, as evidenced by historical contexts such as the emergence of the steam engine and, most notably, nuclear energy. In the 1950s, Günter Anders referred to this phenomenon as the \emph{antiquatedness of man } (1956, 2018). Today, a comparable apprehension of human \emph{obsolescence} and threat is projected onto the future of AI. Notable theoretical physicists, including Stephen Hawking, have expressed concerns in this regard. Given the esteemed status these scientists hold within the scientific community, their concerns have been elevated to the realm of scientific inquiry.
\\
\\
The assumption that humans will be rendered obsolete by AI is based, in part, on the so-called \emph{singularity hypothesis}. This hypothesis likely originated with the British statistician Irvin John Good (1966), who proposed the possibility of an \emph{intelligence explosion}, defined as the development of an ultra-intelligent machine that subsequently produces even more intelligent machines, and so on. The concept of the ``singularity'' is also attributed to writers Stanislaw Lem and Isaac Asimov. This suggests that our ideas and expectations about AI are fundamentally shaped by fiction. Therefore, it is crucial to consider and analyze the cultural and historical components when attempting to clarify the risks that could emanate from AI.
\\
\\
More recently, Ray Kurzweil (2006), Kevin Warwick (2004), and Nick Boström (2014) have argued that the singularity event will play a pivotal role in the development of future AI. The authors present varying ideas and arguments, but they all seek to provide scientific support for the singularity hypothesis.
The singularity seems to be a logical conclusion from the premise of increasing computing power, which is partly based on Moore's Law (1965). However, this law primarily describes the evolution of hardware, i.e. the increase in computing speed in relation to the decrease in the cost of the necessary storage devices. This law has been historically (i.e., empirically) confirmed and has been valid for 60 years. However, it is not a fundamental law of nature, but an economic-technical regularity. Whether this law is valid ``forever'' is at least questionable. In general, the actual progress in AI research is only partly based on the purely quantitative increase in computing power; it also requires the invention of algorithms, the further development of statistics and mathematics, as well as progress in quantum physics. Proponents of the singularity hypothesis also point to the quantum computer, which is expected to provide a gigantic increase in computing power. But the efficiency of modern computing, while necessary for modern learning algorithms and big data, does not seem to be a sufficient condition for the AI singularity.

\subsection{Underestimating the Risks of AI}
In addition to the acknowledged risks of future AI, there is also a lack of awareness regarding the dangers posed by AI that is already in use. Today, certain applications of AI present significant risks to democratic societies, risks that are often underestimated by the majority of people. For instance, the dangers of AI-based f\emph{acial recognition}, \emph{predictive justice}, and \emph{predictive insurance} are frequently overlooked. The concern extends beyond the potential for authoritarian or totalitarian misuse and the susceptibility of humans to \emph{biases}. More fundamentally, it involves the inherent incomprehensibility of the decision-making criteria used by learning algorithms operating with large amounts of data.
\\
\\
\emph{Explainability} has therefore become one of the most important requirements for the application of AI-supported decisions, alongside fairness. Explainability and fairness are essential for trust and social cohesion in human societies. However, in the field of AI applications, there are intractable challenges in integrating explainability into the design. Even today, programmers and computer scientists may lack a clear understanding of how a learning algorithm arrived at a particular decision. Similarly, integrating fairness into AI design is a challenging endeavor.
\\
\\
Nevertheless, there are also promising ideas in this context that warrant greater promotion in the future. One such idea is the 2020 manifesto ``Reasonable Machines: A Research Manifesto'' by Christoph Benzmüller and Bertram Lomfeld. The manifesto's objective is to implement processes of human-like reasoning (deliberation) in AI. Benzmüller and Lomfeld present a series of fundamental concepts in the form of a modeling toolbox. This manifesto should be subjected to rigorous analysis and, potentially, further development. It also illustrates the extent to which people are willing to go in the construction of human-like AI.

\section{Exploring epistemic challenges}
In recent decades, \emph{epistemological} questions in \emph{normative} ethics have seen increasing discussion. For instance, how is moral or normative knowledge constituted and structured? What are the necessary and sufficient conditions for a statement to be morally valid? (e.g., Nida-Rümelin 2020) In this context, it is argued (and not implausibly) that one's beliefs and convictions significantly influence their moral motives for action (cf. e.g., Nida-Rümelin 2020, 335).
\\
\\
For example, it is a known fact that Facebook has sold users' data to third parties without anonymizing it (fact or knowledge). This argues against continuing to use Facebook (moral norm), because this makes it possible to manipulate and instrumentalize people (normative conclusion), and it is generally true that people should not be intentionally instrumentalized (normative or moral criterion). Knowledge of facts and norms is therefore a necessary condition for ethical action. In this example, the relevant knowledge is relatively certain; there are witnesses and leaked documents. But what about data and information of which we can never be sure whether they represent real facts, but which can be highly ethically relevant?

\subsection{The end of explainability?}
The interconnection between knowledge and ethics is becoming increasingly significant, particularly in the context of AI. This is due to the vast amounts of information and data generated by AI or deep learning, which could become morally or ethically relevant at any time. A new discipline has emerged: computational data science (Denning 2017). The ethical challenge posed by the generation of immense volumes of statistical data in the form of correlations, often without corresponding explanations through causality, or sometimes without any possibility of explanation at all, is significant and should not be underestimated. It is evident that the traditional concept of 'knowledge through causalities' is being challenged by petabytes of data. This necessitates reflection on the meaning of causality and the possibility of understanding without a causal explanation. Could there be a crisis in the prevailing notions of causality, knowledge, explanation, and understanding? And what implications would such a crisis have for ethics in general and for AI ethics in particular?
\\
\\
It will be necessary to gain clarity on the problem here so that the questions of AI ethics can be formulated more adequately. It is not excluded that Computational Data Science will help shape the specific ethical questions by opening up the possibilities of new, purely statistical ``knowledge'', on the basis of which automatic ``recommendations'' for ``decisions'' are made that affect ethically sensitive areas like healthcare, for example. We will probably no longer be able to understand these recommendations, but we may have to trust them because or if (and precisely if) they statistically lead to a significant improvement in, for example, healthcare for all citizens.

\subsection{Example of problematization: \emph{ethical governor}}
Ron Arkin’s ethical governor (Arkin et al., 2009) exemplifies the epistemological and ethical challenges inherent to this context. Arkin proposes using AI techniques to ``implement'' a theory of just war. The ethical governor, functioning as a control module, is fed with international law, conventions, treaties of war, and rules of engagement for specific operations. The purpose of the ethical governor is to regulate the decision-making processes of war robots to act more ethically than human soldiers, who can often behave inappropriately (become violent without inhibition) under emotional combat stress. The \emph{jus in bello} rules, i.e., the rules for conducting warfare codified in international law, distinguish between soldiers and civilians and protect the civilian population. However, there are two exceptions to these rules: (1) unarmed soldiers may be taken prisoner but must be protected, and (2) civilians participating in hostilities are considered combatants and may be engaged accordingly. Nevertheless, in asymmetric warfare, where soldiers may not wear uniforms, making such distinctions is exceedingly challenging, particularly for humans. The crucial question is: How can it be ensured that a war robot, including the ethical governor, correctly classifies these situations? This centers on the challenge of making the correct distinctions or classifications.
\\
\\
The highly context-dependent categorization or classification of objects or persons based on situational information, which cannot simply be subsumed under an existing scheme, is therefore problematic and fundamental here. A case-by-case classification or contextual assessment precedes a decision according to the coded rules. An ethical AI would therefore (in addition to solving moral dilemmas) also have to make correct judgments in advance, i.e., make the fine but relevant distinctions about real phenomena. It is not entirely unlikely that war robots could at some point be statistically ``better'' soldiers. Should they exist then? At least in the Ukraine war, those who fight or stand on the side of the moralized good and enforce international law by military means or support this fight would probably welcome the robot soldiers. Wouldn't it be better to abolish war completely with the help of AI instead of developing avoidably ``ethical'' robot soldiers?
\\
\\
In certain areas of medicine, intelligent systems are already far superior in context-dependent classification compared to most experts (e.g., in radiology). Moreover, computer scientists anticipate these capabilities expanding into numerous other domains. Does this not imply the emergence of a new ``narcissistic insult'' in humanity? A new psychological paradigm must be implemented to address this concern.

\section{Investigating fundamental normative issues}
It is assumed that an ethics based on epistemological precepts will inevitably require the development of completely new normative approaches in the future. The data-based knowledge of ethical behavior and the motives of individuals generated by CDS will have a significant impact on ethical theory building and may even result in a fundamental change to the discipline of ethics. AI technologies will be used to better classify the pluralism of ethical approaches and the conflicts of values and norms. This will involve first formulating the normative problems more precisely before attempting to create an ethical or moral AI.
\\
\\
However, there are already concepts and experiments suggesting how an AI could be equipped with ethical or moral programs (Berreby et al. 2018, Bringsjord et al. 2006, Ganascia 2007, Wallach et al. 2008, Powers 2006, Etzioni et al. 2017). The behavior of an AI is assumed to be unpredictable, as it develops opaque subprograms through increasingly sophisticated learning algorithms using vast amounts of training data. The machine's behavior then becomes not only incomprehensible to humans but can also harm them. The prevailing view is, therefore, that AI behavior should be (ethically) programmed from the outset to conform to shared social norms and values.
\\
\\
As part of the radical critique, it is necessary to conduct a systematic investigation into the possibilities of programming AI ethically or normatively. This investigation must also consider the fundamental limits and paradoxes that such an endeavor may encounter.

\subsection{Modeling the normative}
Research over the last 50 years has shown that, firstly, modeling deontic and normative-logical reasoning is extremely difficult. Secondly, there is currently no clear method (apart from randomization or arbitrariness) for handling notorious conflicts of norms and moral dilemmas in AI. Thirdly, the link between motivation or principles, decision, and action cannot yet be modeled. This is because there seems to be a consensus that not only the principles of an action but also the assessments of the consequences of this action are ethically relevant.
\\
\\
To address the first challenge, some researchers have drawn inspiration from the formalisms of deontic logic (Lorini 2012, Horty 2001), despite the presence of numerous paradoxes within this framework. The second problem is approached using AI-logic-based formalisms of non-monotonic logic (Ganascia 2015) to emulate rational deliberation, such as default logic (Reiter 1980) or answer set programming (Gelfond 2008). For the third problem, logic-based models of ethical judgment are combined with formalisms of so-called action languages (Mueller 2014) or causal models (Halpern et al. 2018) to achieve clear semantics and thereby establish a mathematical basis for automatically incorporating the consequences of actions.
\\
\\
A closer examination of the three approaches mentioned above reveals that they correspond to the three different paradigms of normative ethics: the first to deontology, the second to virtue ethics, and the third to consequentialism. Despite their abstract nature, these approaches are not normatively neutral. Moreover, when considered together, they can contradict each other, similar to the three normative ethical theories. This inconsistency arises because normative ethics lacks a unified and consistent theory of the normative. Although these formal approaches may be highly complex, they often fail to adequately depict ethical behavior. Therefore, the fundamental issue lies not solely in the formalizations or their inherent problems (which undoubtedly exist), but rather in the broader challenge of ethics or the theory of normativity itself.
\\
\\
In the context of scientifically guided critique, these various approaches aiming to formally encapsulate different aspects of ethics should be thoroughly examined and critically questioned. Should we continue attempting to formalize traditional moral philosophical concepts and theories, given their failure to solve theoretical or practical problems?
\\
\\
The hypothesis here is that the disparate formal languages and approaches used to formalize moral philosophy and normative ethics cannot provide a stable foundation for a consistent normative theory, and thus cannot effectively guide the normative programming of AI. Each of these approaches is underdetermined in normative content individually, and they become either overdetermined or inconsistent when combined. The fundamental problem of normativity remains unresolved to this day. Traditional moral philosophical and ethical theories, no matter how sophisticated their formal disguises today, are at an impasse. What is urgently needed is a new science, a science of normativity.

\subsection{Problematization: \emph{moral machine}}
There are already several machine learning projects exploring how AI could automatically learn values and norms (Abel et al. 2016, Kleiman-Weiner 2017). The popularity and effectiveness of machine learning are driving these initiatives. However, philosophers often criticize these projects, arguing that ethics should not be based on data—specifically, not on empirical observation and automated classification of observable human behavior.
\\
\\
The widely discussed online platform ``Moral Machine''\f{$https://www.moralmachine.net/$} serves as an illustrative example of this approach (Awad et al. 2018). In this experiment, researchers collect and analyze individuals’ decision-making processes in response to various car crash scenarios designed as moral dilemmas. Data from test subjects or users is categorized based on parameters such as region, culture, religion, and gender, revealing significant variations in moral attitudes and principles. While this experiment provides valuable insights for social and moral psychology, its primary objective is not scientific comparisons between cultures. Instead, the platform aims to determine how self-driving cars, in particular, could be programmed in a “moral” manner that aligns with local culture and the expectations of its presumably homogeneous clientele.
\\
\\
The online experiment touches upon many controversies within philosophical ethics. However, the experimenters and programmers are partly aware of these controversies and argue that traditional methods used by ethicists and philosophers to address moral dilemmas and conflicts of norms have often proven ineffective in practical applications. They do not necessarily advocate for the universal application of values and norms observed within a surveyed society or culture. The experimenters acknowledge and reflect on this distinction (Awad et al. 2018). Thus, there is an inherent issue here with interdisciplinary collaboration, and a radical clarification is necessary to illuminate the relationship between computer scientists, scientists, and philosophers.
\\
\\
From the perspective of the ``moral machine'' concept, a sort of ``alienation'' from traditional philosophical ethics becomes apparent. Embracing this empirical, data-based approach would imply moving away from philosophical ethics, at least partially, and leveraging the advantages of machine learning in studying moral and normative behavior.

\section{Outlook}
The most pressing questions surrounding AI are inherently normative, and akin to the fundamental queries of normativity itself, they remain unresolved both theoretically and practically. Currently, normative or moral judgments are predominantly made in a traditional, everyday manner, guided by vague criteria and intuitive considerations. Proposals from normative ethics serve as loose guidelines in AI applications and regulations, often drawing upon deontological, consequentialist, or virtue ethical reasoning. However, these ethical frameworks are inherently underdetermined and can contradict each other in hybrid forms, posing challenges for their implementation in AI ethics. Consequently, they cannot be readily implemented in AI (ethics).
\\
\\
It seems unlikely that a coherent AI ethics will emerge without a deep re-evaluation of the fundamental concepts and theories within normative ethics. The development of a potential new science of the normative is necessary here. Although it may appear to be a circular argument, the development of new theoretical approaches to normativity is likely to depend on advances in AI. The radical questioning of the fundamental concepts and theories of normative ethics, is a prerequisite for developing a new science of normativity. This, in turn, is a necessary condition for creating a contradiction-free ethical framework for artificial intelligence.
\\
\\
\\
\\
\begin{large}{\textbf{Bibliography}}
\end{large}
\begin{description}
\item[]Abel, D. et al. (2016),  Reinforcement Learning as a Framework for Ethical Decision Making, Palo Alto, CA: \emph{Association for the Advancement of Artificial Intelligence}, 2016): 54–61.
\item[]Arkin, R. et al. (2009), An Ethical Governor for Constraining Lethal Action in an Autonomous System, \emph{Technical Report GIT-GVU-09-02}, Georgia Institute of Technology Mobile Robot Lab.
\item[]Asimov, I. (1956), The Last Question, \emph{Science Fiction Quarterly} 4:5 (Nov.).
\item[]Awad, E. et al. (2018), The Moral Machine Experiment, \emph{Nature} 563: 59–64.
\item[]Benzmüller, Ch., Lomfeld, B. (2020), Reasonable Machines: A Research Manifesto, \emph{ArXiv:} https://arxiv.org/abs/2008.06250.
\item[]Bringsjord, S. et al.  (2006),  Toward a General Logicist Methodology for Engineering Ethically Correct Robots, \emph{IEEE Intelligent Systems }21:4: 38–44.
\item[]Bostrom, N. (2014), Superintelligence: Path, Dangers, Strategies, Oxford University Press.
\item[]Chalmers, D.J. (1995), Facing up to the Problem of Consciousness, \emph{Journal of Consciousness Studies} 2: 200–219.
\item[]Dehaene, S. et al. (2017), What Is Consciousness, and Could Machines Have It? \emph{Science} 358:6362: 486–92.
\item[]Dennett, D.C. (1971), Intentional Systems, \emph{Journal of Philosophy} 68:4: 87–106.
\item[]Dennett, D.C. (1987), The Intentional Stance, Cambridge, MA: MIT Press.
\item[]Denning, P. J. (2017), Computational Thinking in Science, \emph{American Scientist} 105:1: 13–17.
\item[]Etzioni, A. \& O. (2017), Incorporating Ethics into Artificial Intelligence, \emph{Journal of Ethics }21:4 (2017): 403–18.
\item[]Gabriel, M. (2018), Der Sinn des Denkens. 1. Auflage. Ullstein, Berlin.
\item[]Ganascia, J.-G. (2007), Modelling Ethical Rules of Lying with Answer Set Programming, \emph{Ethics and Information Technology} 9:1: 39–47.
\item[]Ganascia, J.-G. (2015), Non-monotonic Resolution of Conflicts for Ethical Reasoning, in:  A Construction Manual for Robots’ Ethical Systems, ed. Robert Trappl, Springer International Publishing, 101–18.
\item[]Gelfond, M. (2008), Answer Sets, \emph{Foundations of Artificial Intelligence} 3: 285–316.
\item[]Good, I. J. (1966), Speculations Concerning the First Ultraintelligent Machine, \emph{Advances in Computers} 6: 31–88.
\item[]Harari, Yuval Noah (2018), 21 Lessons for the 21st Century, 1. Edition.
\item[]Horty, J. F. (2001) Agency and Deontic Logic, Oxford: Oxford University Press.
\item[]Kleiman-Weiner, M. et al. (2017), Learning a Common-Sense Moral Theory, \emph{Cognition }167: 107–23.
\item[]Kurzweil, R. (2006), The Singularity Is Near: When Humans Transcend Biology, New York: Penguin Books, 2006.
\item[]Lorini, E. (2012),  On the Logical Foundations of Moral Agency, \emph{Lecture Notes in Computer Science }7393, Berlin: Springer, 108–22.
\item[]McCarthy, J. et al. (1955), A Proposal for the Dartmouth Summer Research Project on Artificial Intelligence, \emph{AI Magazine}, 27(4), 12. https://doi.org/10.1609/aimag.v27i4.1904.
\item[]Misselhorn, C. (20018), Grundfragen der Maschinenethik, 3., durchges. Aufl., Stuttgart.
\item[]Mueller, E. T. (2014), Commonsense Reasoning: An Event Calculus Based Approach (Burlington, MA: Morgan Kaufmann).
\item[]Newell, A./Simon, H.A. (1972), Human Problem Solving, Englewood Cliffs, New Jork: Prentice-Hall.
\item[]Nagel, Thomas (1974), What Is It Like to Be a Bat?. \emph{The Philosophical Review} 83 (4): 435–450.
\item[]Nida-Rümelin, J. (2018), Digitaler Humanismus, Piper, München.
\item[]Nida-Rümelin, J. (2020), Eine Theorie der praktischen Vernunft, De Gruyter, Berlin.
\item[]O’Neil, C. (2016), Weapons of Math Destruction, New York: Crown Publishers.
\item[]Powers, Th. et al. (2006), Prospects for a Kantian Machine, \emph{IEEE Intelligent Systems} 21:4 (2006): 46–51.
\item[]Reiter, R. (1980), A Logic for Default Reasoning, \emph{Artificial intelligence} 13:1–2: 81–132.
\item[]Russell, S. J./Norvig, P. (2010), Artificial Intelligence: A Modern Approach, 3rd ed., Upper Saddle River, NJ: Prentice-Hall.
\item[]Wallach, W. et al. (2008), Machine Morality: Bottom-up and Top-down Approaches for Modelling Human Moral Faculties, \emph{AI \& Society} 22:4: 565–82.
\item[]Warwick, K. (2004),  March of the Machines: The Breakthrough in Artificial Intelligence, Champaign: University of Illinois Press.
\end{description}

\newpage

\end{document}